\documentclass[submitting]{NST}

\usepackage{subfigure,dcolumn}
\usepackage[T2A,T1]{fontenc}
\usepackage[russian,english]{babel}
\usepackage{listings}
\begin{document}

\title{Effects of sequential decay on collective flows and nuclear stopping power in heavy-ion collisions at intermediate energies}

	\author{Kui Xiao}
	\affiliation{State Key Laboratory of Quantum Optics and Quantum Optics Devices and Collaborative Innovation Center of Extreme Optics, Institute of Theoretical Physics, Shanxi University, Taiyuan 030006, China}
	\affiliation{School of Science, Huzhou University, Huzhou 313000, China}
	\author{Pengcheng Li}
	\affiliation{School of Science, Huzhou University, Huzhou 313000, China}
	\affiliation{School of Nuclear Science and Technology, Lanzhou University, Lanzhou 730000, China}
	\affiliation{Institut f\"{u}r Theoretische Physik, Goethe Universit\"{a}t Frankfurt, Frankfurt am Main 60438, Germany}
	\author{Yongjia Wang}
	\email[Corresponding author, ]{wangyongjia@zjhu.edu.cn}
	\affiliation{School of Science, Huzhou University, Huzhou 313000, China}
	\author{Fuhu Liu}
	\affiliation{State Key Laboratory of Quantum Optics and Quantum Optics Devices and Collaborative Innovation Center of Extreme Optics, Institute of Theoretical Physics, Shanxi University, Taiyuan 030006, China}
	\author{Qingfeng Li}
	\email[Corresponding author, ]{liqf@zjhu.edu.cn}
	\affiliation{School of Science, Huzhou University, Huzhou 313000, China}
	\affiliation{Institute of Modern Physics, Chinese Academy of Science, Lanzhou 730000, China}

\begin{abstract}
In this study, the rapidity distribution, collective flows, and nuclear stopping power in $^{197}\mathrm{Au}+^{197}\mathrm{Au}$ collisions at intermediate energies were investigated using the ultrarelativistic quantum molecular dynamics (UrQMD) model with GEMINI++ code. The UrQMD model was adopted to simulate the dynamic evolution of heavy-ion collisions, whereas the GEMINI++ code was used to simulate the decay of primary fragments produced by UrQMD. The calculated results were compared with the INDRA and FOPI experimental data. It was found that the rapidity distribution, collective flows, and nuclear stopping power were affected to a certain extent by the decay of primary fragments, especially at lower beam energies. Furthermore, the experimental data of the collective flows and nuclear stopping power at the investigated beam energies were better reproduced when the sequential decay effect was included.
\end{abstract}


\maketitle

\section{Introduction}
\label{sec:1}
Understanding the properties of dense nuclear matter is a main topic of the basic research on heavy-ion collisions (HICs), as it is the only way of creating dense nuclear matter in terrestrial laboratories \cite{Bertsch:1988ik,Aichelin:1991xy,Danielewicz:2002pu,li2008recent,xu2019transport}. However, the typical time scale for the appearance of the created dense nuclear matter is extremely short (less than or of the order of a few fm/$c$) \cite{Guo:2022kwc}; thus, direct measurement of its properties is currently impossible. Usually, the fundamental properties of dense nuclear matter are inferred by comparing experimental data and transport model simulations \cite{Ono:2019jxm,bleicher2022modelling,colonna2020collision,huth2022constraining,Lan:2022rrc}. As powerful tools for investigating the dynamics of nonequilibrium systems at intermediate energies, two typical transport models, the Boltzmann–Uehling–Uhlenbeck (BUU)-type \cite{Bertsch:1988ik} and quantum molecular dynamics (QMD)-type models \cite{Aichelin:1991xy} and their updated versions, have been extensively used for simulating HICs to extract the structure of the initial nuclei and the properties of the created dense nuclear matter, such as the nuclear equation of state, the nuclear symmetry energy, and the in-medium nucleon-nucleon cross section \cite{Shi:2021far,Lin2021,Wang:2021xpv,Wang:2021jgu,Liu:2022ewr,Wang:2022hyi}. 
	
Both the BUU- and QMD-type models mainly include three components: the initialization of the projectile and target nuclei, mean-field potential for each particle, and two-body scattering (collision term). Different techniques are typically used to treat the three components in different models. To establish a theoretical systematic error that quantifies the model dependence of transport predictions and disentangle the causes of different predictions, the transport model evaluation project (TMEP) has been recently performed \cite{Xu:2016lue,TMEP:2022xjg,Colonna:2021xuh,Ono:2019ndq,Zhang:2017esm}. 
After simulating the dynamical process of HICs within the transport models, an afterburner is usually chosen for the formation of fragments, which can then be used to construct observables, and further compared with the corresponding experimental data to extract interesting information regarding the nuclear matter. However, the formed fragments are typically excited and cannot be experimentally detected. In the experiments, the detected particles underwent long-term de-excitation. Hence, greater focus should be placed on the effect of de-excitation in the treatment of afterburners, especially for studying HICs at Fermi energies where the beam energy is comparable to the exciting energy \cite{Tian:2006zr,Su:2018lco,1835162}. 
	
The underestimation of the yield of light fragments (e.g., $^3$H, $^3$He and $^4$He) in HICs at intermediate energies is a long-standing problem in transport model simulations, and numerous techniques have been developed and adopted to overcome it \cite{Tsang:2006zc}, including the statistical multifragmentation model (SMM) \cite{Bondorf:1985mv,Bondorf:1995ua}, the statistical evaporation model (HIVAP) \cite{reisdorf1981analysis}, the statistical model GEMINI \cite{Charity:1988zz}, and the Simulated Annealing Clusterization Algorithm (SACA) \cite{Puri:1996qv,Puri:1998te}. 
From our previous studies \cite{Wang:2013wca,Li:2018bus,Li:2022wvu}, adopting the conventional phase space coalescence model “Minimum Spanning Tree” (MST) \cite{Li:2016wkb,Li:2016oxs,Li:2016mqd}and disregarding the decay of the excited primary fragments from the Ultra-relativistic Quantum Molecular Dynamics (UrQMD) model simulation, the experimental data of the collective flow of protons and deuterons at INDRA and FOPI energies can be reasonably reproduced, whereas those of $\alpha$ particles are different from the experimental data. To improve the corresponding collective flow distribution and explore the influence of sequential decay on the observables in the HICs at intermediate energies, the statistical code GEMINI++ was employed to describe the decay of primary fragments.

In this study, the rapidity distributions and collective flows of free protons and light clusters, as well as the nuclear stopping power, were calculated based on the UrQMD model with and without considering sequential decay. The remainder of this paper is organized as follows. In Sect. \ref{sec:2}, the relevant descriptions of the transport model, statistical model, and the observables are provided. The calculation results are presented and discussed in Sect. \ref{sec:3}. Finally, a summary and outlook are provided in Sect. \ref{sec:4}.
%%%%%%%%%%%%%%%%%%%%%%%%%%%%%%%%%%%%%%%%%%%%%%%%%%%%%%%%
\section{Model description and observables}
\label{sec:2}
%%%%%%%%%%%%%%%%%%%%%%%%%%%%%%%%%%%%%%%%%%%%%%%%%%%%%%%%	
\subsection{Dynamical model: UrQMD}
The UrQMD model \cite{Bass:1998ca,Bleicher:1999xi,Li:2011zzp,Wang:2020vwb} is a typical transport model used for microscopic many-body nonequilibrium dynamics. In this model, each nucleon is represented by a Gaussian wave packet of a certain width $L$. Empirically, $L$ = 2 fm$^2$ are chosen to simulate Au+Au collisions. The coordinates and momentum of each nucleon are propagated using Hamilton’s equations of motion. The total Hamiltonian of the system $\langle H \rangle$ consists of the kinetic energy $T$ and the effective interaction potential energy $U$, which includes the Skyrme potential energy $U_{\rho}$, Coulomb energy $U_{Coul}$ and momentum dependent potential energy $U_{md}$:
	\begin{equation}
		\langle H \rangle=T+U_{\rho}+U_{Coul}+U_{md}.
	\end{equation}
To study the HICs at intermediate energies, the Skyrme energy density functional was introduced in the same manner as that in the improved quantum molecular dynamics (ImQMD) model\cite{Zhang:2020dvn,Wang:2014aba}. The local and momentum-dependent potential energies can be written as $U_{\rho,md}=\int{u_{\rho,md}d\textbf{r}}$ where
	\begin{equation}\label{urho}
		\begin{aligned}
			u_{\rho}=
			&\frac{\alpha}{2}\frac{\rho^2}{\rho_{0}}+\frac{\beta}{\gamma+1}\frac{\rho^{\gamma+1}}{\rho_{0}^{\gamma}}+\frac{g_{\text{sur}}}{2\rho_{0}}(\nabla\rho)^{2}\\
			&+\frac{g_{\text{sur,iso}}}{2\rho_{0}}[ \nabla(\rho_{n}-\rho_{p})] ^{2}+\left [ A_{\text{sym}}(\frac{\rho}{\rho_{0}}) \right.\\ 
			&\left.+B_{\text{sym}}(\frac{\rho}{\rho_{0}})^{\eta}+C_{\text{sym}}(\frac{\rho}{\rho_{0}})^{5/3}\right ]\delta^2\rho,
		\end{aligned}
	\end{equation}
	and
	\begin{equation}
		u_{md}=t_{md} \ln^{2}\left[1+a_{md} (\textbf{p}_{i}-\textbf{p}_{j})^2\right] \frac{\rho}{\rho_{0}}.
	\end{equation}
Here, $\delta=\left(\rho_{n}-\rho_{p}\right) /\left(\rho_{n}+\rho_{p}\right)$ denotes the isospin asymmetry defined by the neutron($\rho_{n}$) and proton ($\rho_{p}$) densities.
In the present work, the normal nuclear matter density $\rho_{0}$ = 0.16 fm$^{-3}$, $\alpha=-$393 MeV, $\beta$ = 320 MeV, $\gamma$ = 1.14, $g_{\text{sur}}$ = 19.5 MeV fm$^{2}$, $g_{\text{sur,iso}}=-$11.3 MeV fm$^{2}$, $A_{\text{sym}}$ = 20.4 MeV, $B_{\text{sym}}$ = 10.8 MeV, $C_{\text{sym}}=-$9.3 MeV, $\eta$ = 1.3 , $t_{md}$ = 1.57 MeV, and $a_{md}$=500 (GeV/$c$)$^{-2}$ are chosen, corresponding to a soft and momentum-dependent (SM) equation of state with the incompressibility $K_{0}$ = 200 MeV and the symmetry-energy slope parameter $L$ = 80.95 MeV, which are in their commonly accepted regions \cite{Wang:2020vwb,Wang:2018hsw,Xu:2022squ,Wei:2021arw,Liu:2021uoz,wang2020study}.
	
In this work, the in-medium nucleon-nucleon (NN) elastic cross section is considered as the result of a density- and momentum-dependent medium correction factor $\mathcal{F}(\rho, p)$ times the free nucleon-nucleon cross section $\sigma_{N N}^{\mathrm{free}}$, which is available from the experimental data.
		\begin{equation}
			\sigma_{N N}^{\mathrm{in}-\mathrm{medium}}=\mathcal{F}(\rho, p) * \sigma_{N N}^{\mathrm{free}},
		\end{equation}
		with
		\begin{equation}
			\mathcal{F}(\rho,p)=\left\{
			\begin{array}{l}
				f_0, \hspace{3.1cm} p_{NN}>1 {\rm GeV}/c, \\
				\frac{\lambda+(1-\lambda)e^{-\frac{\rho}{\zeta\rho_{0}}} -f_{0}}{1+(p_{NN}/p_{0})^\kappa}+f_{0}, \hspace{0.2cm} p_{NN} \leq 1 {\rm GeV}/c.
			\end{array}
			\right.
			\label{fdpup}
		\end{equation}
The parameters $f_{0}$ = 1, $\lambda$ = 1/6, $\zeta$ = 1/3, $p_{0}$ = 0.3 GeV/$c$ and $\kappa$ = 8 were adopted, with $p_{NN}$ being the momentum in the two-nucleon center-of-mass frame. More detailed studies regarding the effects of $\mathcal{F}(\rho, p)$ parameters on various observables can be found in Refs.~\cite{Li:2011zzp,Wang:2020vwb,Li:2018wpv}. In addition to the in-medium $NN$ cross section, Pauli blocking plays an important role in determining the rate of collisions. The Pauli blocking treatment is the same as that described in Refs.~\cite{Li:2011zzp,TMEP:2022xjg}; it involves two steps. First, the phase space densities $f_i$ and $f_j$ of the two outgoing particles for each $NN$ collision are calculated.
\begin{equation}
f_{i} = \frac{1}{2}\sum_k  \frac{1}{(\pi\hbar)^3}\,\exp\big[-\frac{(\textbf{r}_i-\textbf{r}_k)^2}{(2\sigma^2_r)}\bigr] \, \exp\big[-\frac{(\textbf{p}_i-\textbf{p}_k)^2\cdot2\sigma^2_ r}{\hbar^2}\bigr]
 \label{eq5urqmd}.
\end{equation}
Here, $\sigma^2_ r$ denotes the width parameter of the wave packet. $\sigma^2_r$ = 2 fm$^{2}$ is used in this work. $k$ represents nucleons with the same type around the outgoing nucleon $i$ (or $j$). The following two criteria were simultaneously considered:
\begin{equation}
\frac{4\pi}{3}r_{ik}^3\frac{4\pi}{3}p_{ik}^3 \geq 2\left(\frac{h}{2}\right)^3
 \label{eq6urqmd};
\end{equation}

\begin{equation}
P_{block}=1-(1-f_i)(1-f_j) <\xi
 \label{eq7urqmd}.
\end{equation}
Here, $r_{ik}$ and $p_{ik}$ denote the relative distance and momentum between nucleons $i$ and $k$. The above conditions are also considered for nucleon $j$. The symbol $\xi$ denotes a random number between 0 and 1. 

In addition, the isospin-dependent minimum spanning tree (iso-MST) algorithm was used to construct the clusters \cite{Li:2016wkb,Li:2016oxs,Li:2016mqd}. Nucleon pairs with relative distances smaller than $R_{0}$ and relative momenta smaller than $P_{0}$ = 0.25 GeV/$c$ are considered to be bound in a fragment. Here, $R_{0}^{pp}$ = 2.8 fm and $R_{0}^{np}=R_{0}^{nn}$ = 3.8 fm are used for proton-proton and neutron-neutron (neutron-proton) pairs, respectively. A fairly good agreement between the recently published experimental data (including collective flows, nuclear stopping power, and Hanbury–Brown–Twiss interferometry) and UrQMD model calculations was achieved with appropriate choices of the above parameters \cite{Wang:2018hsw,Li:2022wvu,Wang:2020vwb,lipc1,lipc2}.
%%%%%%%%%%%%%%%%%%%%%%%%%%%%%%%%%%%%%%%%%%%%%%%%%%%%
\subsection{Statistical model: GEMINI++}
The statistical GEMINI++ decay model \cite{Charity:1988zz,Charity:2010wk} is a Monte Carlo method to simulate the decay of excited fragment, which includes light particle evaporation, symmetric and asymmetric fission, and all possible binary-decay modes. It has been widely used to treat the de-excitation of the primary fragments in studies of nuclear reactions at low and intermediate energies \cite{Hagel:1994zz,Tian:2006zr,Su:2011zze,Dai:2015dua,Ma:2002ix,Huang:2010jp,LiLi:2022kvc,1835162,Su:2018veq} and high energies \cite{Liu:2022xlm,Lei:2023onu,Song:2022yjr}. The parameters of the shell-smoothed level density in GEMINI++ were set to their default values, which were $k_{0}$ = 7.3 MeV and $k_{\infty }$ = 12 MeV. A more detailed discussion of the GEMINI++ parameters can be found in Refs.~\cite{Su:2018veq,Charity:2010wk,Mancusi:2010tg}.
  
There are four inputs for the GEMINI++ code (with default parameter settings): the mass number $A$, charge number $Z$, excitation energy $E^{*}$, and angular moment $\vec{L}$ of the primary fragment. After the UrQMD transport and iso-MST coalescence processes, the excitation energy of a fragment is calculated as follows:
		\begin{equation}
			E^{*}=E_{\rm{bind}}^{\rm{excited}}-E_{\rm{bind}}^{\rm{ground}},
		\end{equation}
where $E_{\rm{bind}}^{\rm{excited}}$ is the binding energy of the primary fragment, that is, the sum of the potential and kinetic energies of all nucleons that belong to a fragment. $E_{\rm{bind}}^{\rm{ground}}$ is the binding energy of the ground state obtained from nuclear data table AME2020 \cite{Wang:2021xhn,gaozp}. When the excitation energy of the fragment was less than or equal to zero, it was regarded as a bound state and did not decay further. The angular momentum of the primary fragments was calculated using classical mechanics \cite{Dai:2015dua}.
		\begin{equation}
			\vec{L}=\sum_{i} \vec{r}_{i} \times \vec{p}_{i},
		\end{equation}
where $\vec{r}_{i}$ and  $\vec{p}_{i}$ are the coordinate and momentum vectors, respectively, of the $i$-th nucleon in the primary fragments in the c.m. frame of the fragment. The total angular momentum is the sum of all the nucleons in the primary fragments. We verified that the results remained almost unchanged if the contribution of the angular momentum in GEMINI++ was not included.
%%%%%%%%%%%%%%%%%%%%%%%%%%%%%%%%%%%%%%%%%%%%%%%%%%%%%%	
\subsection{Observables}
In this study, we focus on collective flows and nuclear stopping power, which are commonly used observables, and propose investigating the properties of dense nuclear matter. The directed ($v_{1}$) and elliptic ($v_{2}$) flows are defined as \cite{Reisdorf:1997fx}
		\begin{equation}
			v_{1}=\left\langle\frac{p_{x}}{\sqrt{p_{x}^2+p_{y}^2}}\right\rangle,~~~~v_{2}=\left\langle\frac{p_{x}^{2}-p_{y}^{2}}{\sqrt{p_{x}^2+p_{y}^2}}\right\rangle.
		\end{equation}	
Here, $p_{x}$, $p_{y}$ are the two components of the transverse momentum $p_{t}=\sqrt{p_{x}^{2}+p_{y}^{2}}$. Angle brackets indicate the average of all considered particles from all events. The nuclear stopping power $R_{E}$, which characterizes the transparency of the colliding nuclei, can be defined as \cite{INDRA:2010pbz},
		\begin{equation}
			R_{E}=\frac{\sum E_{\perp}}{2 \sum E_{\|}},
		\end{equation}
where $E_{\perp}$ ($E_{\|}$) is the c.m. transverse (parallel) energy, and the sum runs over all considered particles. Another quantity $vartl$ is also widely used to measure the stopping power, which was proposed by the FOPI Collaboration \cite{FOPI:2011aa}, is widely used to measure stopping power. It is defined as the ratio of the variances of the transverse to those of the longitudinal rapidity distribution and is expressed as
		\begin{equation}
			vartl=\frac{\left\langle y_{x}^{2}\right\rangle}{\left\langle y_{z}^{2}\right\rangle}.
		\end{equation}
where $\left\langle y_{x}^{2}\right\rangle$ and $\left\langle y_{z}^{2}\right\rangle$ are the variances in the rapidity distributions of the particles in the $x$ and $z$ directions, respectively. 
%%%%%%%%%%%%%%%%%%%%%%%%%%%%%%%%%%%%%%%%%%%%%%%%%%%
\section{Results and discussions}
\label{sec:3}
%%%%%%%%%%%%%%%%%%%%%%%%%%%%%%%%%%%%%%%%%%%%%%%%%%%
\subsection{Rapidity distribution}

		\begin{figure}[b]
			\centering
			\includegraphics[width=\linewidth]{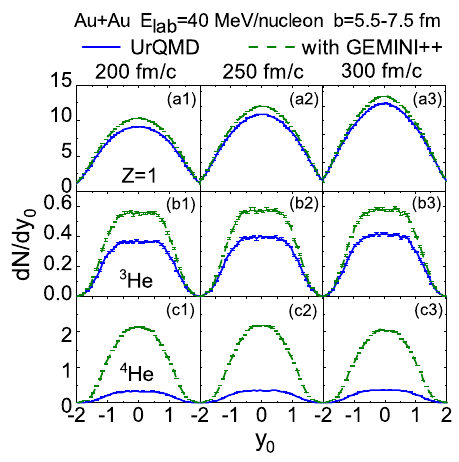}
			\caption{Rapidity distributions of Z = 1 [panels (a1-a3)], $^3\mathrm{He}$ [panels (b1-b3)] and $^4\mathrm{He}$ [panels (c1-c3)] particles as functions of the reduced rapidities($y_0=y_z/y_{pro}$) from peripheral ($b$ = 5.5-7.5 fm) $^{197}\mathrm{Au}+^{197}\mathrm{Au}$ collisions at $E_{\rm{lab}}$ = 40 MeV/nucleon at different reaction times. The solid and dashed lines represent the results obtained from UrQMD simulations without and with considering the decay of the primary fragments by using GEMINI++, respectively}
			\label{fig:1}
		\end{figure}
  
The rapidity distributions of the hydrogen (Z = 1) and helium isotopes ($^{3}$He and $^{4}$He) in $^{197}\mathrm{Au}+^{197}\mathrm{Au}$ collisions at $E_{\rm{lab}}$ = 40 MeV/nucleon and impact parameter $b$ = 5.5-7.5 fm are shown in Fig.~\ref{fig:1}. The left, middle and right panels are the calculated results when the switching time of UrQMD model was set to 200 fm/$c$, 250 fm/$c$ and 300 fm/$c$, respectively. When considering the decay of the excited primary fragments, the yields of Z = 1 particles are slightly increased, whereas those of $^3$He and $^4$He clusters are clearly increased. The yields of $^3$He and $^4$He are approximately 1.5 and 6 times greater than those obtained from the simulations without considering sequential decay. It is known that the yields of light clusters such as $^3$He and $^4$He obtained using QMD-type codes are significantly smaller than experimentally measured yields. As discussed in our previous work \cite{Wang:2014aba}, the yield of $^3$He calculated using the UrQMD model was approximately three times lower than that of the experimental data. The inclusion of GEMINI++ improved the description of light cluster production to some extent; however, the yields of light clusters were still underestimated. Additionally, the inclusion of GEMINI++ may have led to an overestimation of the yield of free nucleons. To solve these problems further, additional issues should be considered, such as the spin degree of freedom and the dynamical production process \cite{Ono:2019jxm}.
		
%%%%%%%%%%%%%%%%%%%%%%%%%%%%%%%%%%%%%%%%%%%%%%%%%%
\subsection{Collective flows}
  
		\begin{figure}[t]
			\centering
			\includegraphics[width=\linewidth]{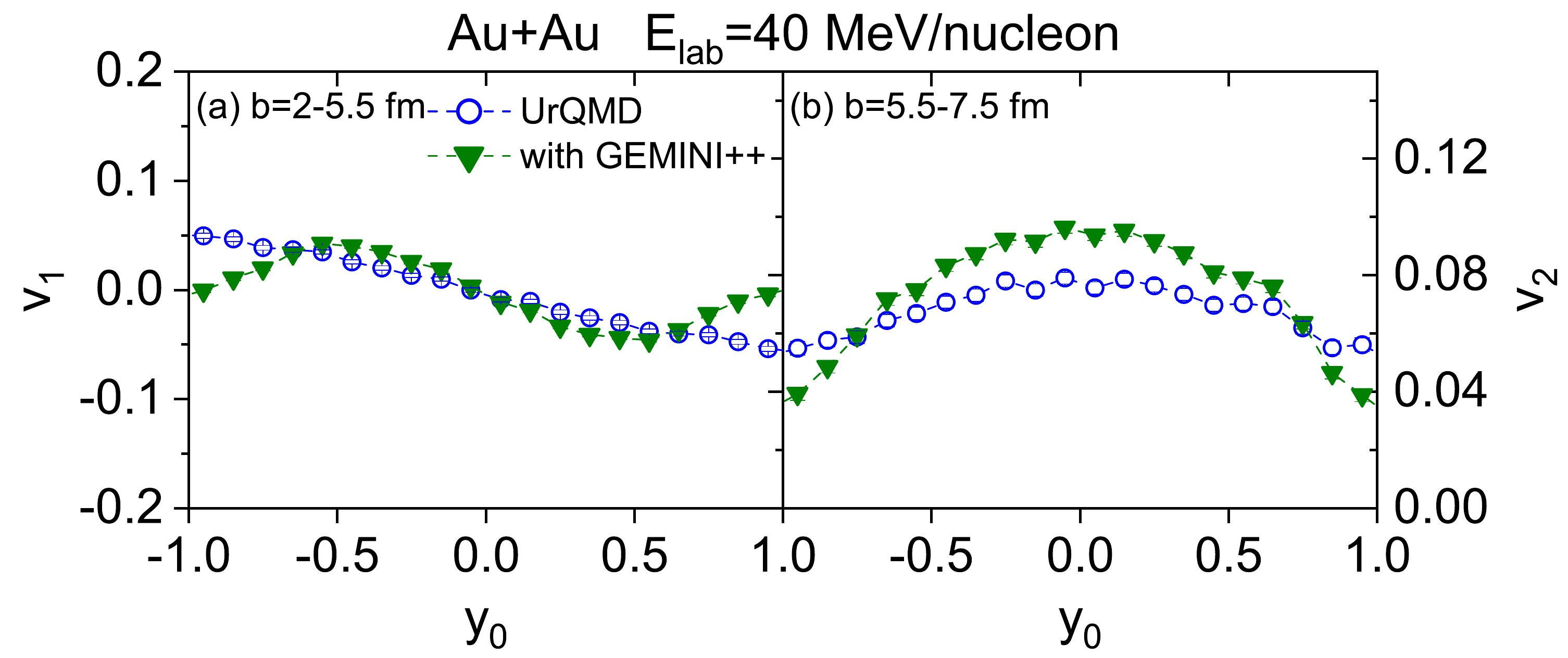}
			\caption{Reduced rapidity $y_{0}$ distributions of the directed (elliptic) flow $v_{1}$ ($v_{2}$) from semi-central (peripheral) $^{197}\mathrm{Au}+^{197}\mathrm{Au}$ collisions at $E_{\rm{lab}}$ = 40 MeV/nucleon for Z = 1 particles. The blue circles denote the calculated results without primary fragments decay, whereas the olive down-triangles represent the calculated results with primary fragments de-excited}
			\label{fig:2}
		\end{figure}
		
The collective flows of Z = 1 particles as functions of $y_{0}$ in $^{197}\mathrm{Au}+^{197}\mathrm{Au}$ collisions at $E_{\rm{lab}}$ = 40 MeV/nucleon are shown in Fig.~\ref{fig:2} (directed flow $v_{1}$ [left panel] and elliptic flow $v_{2}$ [right panel]). The simulations are stopped at 200 fm/$c$. Clear differences are observed in the distributions of the collective flows between simulations with and without considering the decay of the primary fragments, especially in the case of elliptic flow $v_{2}$ from peripheral ($b$ = 5.5-7.5 fm) collisions. From Fig.~\ref{fig:1} the difference in the yields of Z = 1 particles between simulations with and without considering sequential decay is due to the particles produced from the decay of excited primary fragments, which have larger mass. These heavier excited primary fragments are usually produced at the target/projectile rapidity; thus, the directed flow at the target/projectile rapidity is significantly affected by sequential decay. Increasing the impact parameter results in more large-mass primary fragments being produced and the effects of sequential decay on the collective flows becoming more evident.

		\begin{figure}[t]
			\centering
			\includegraphics[width=\linewidth]{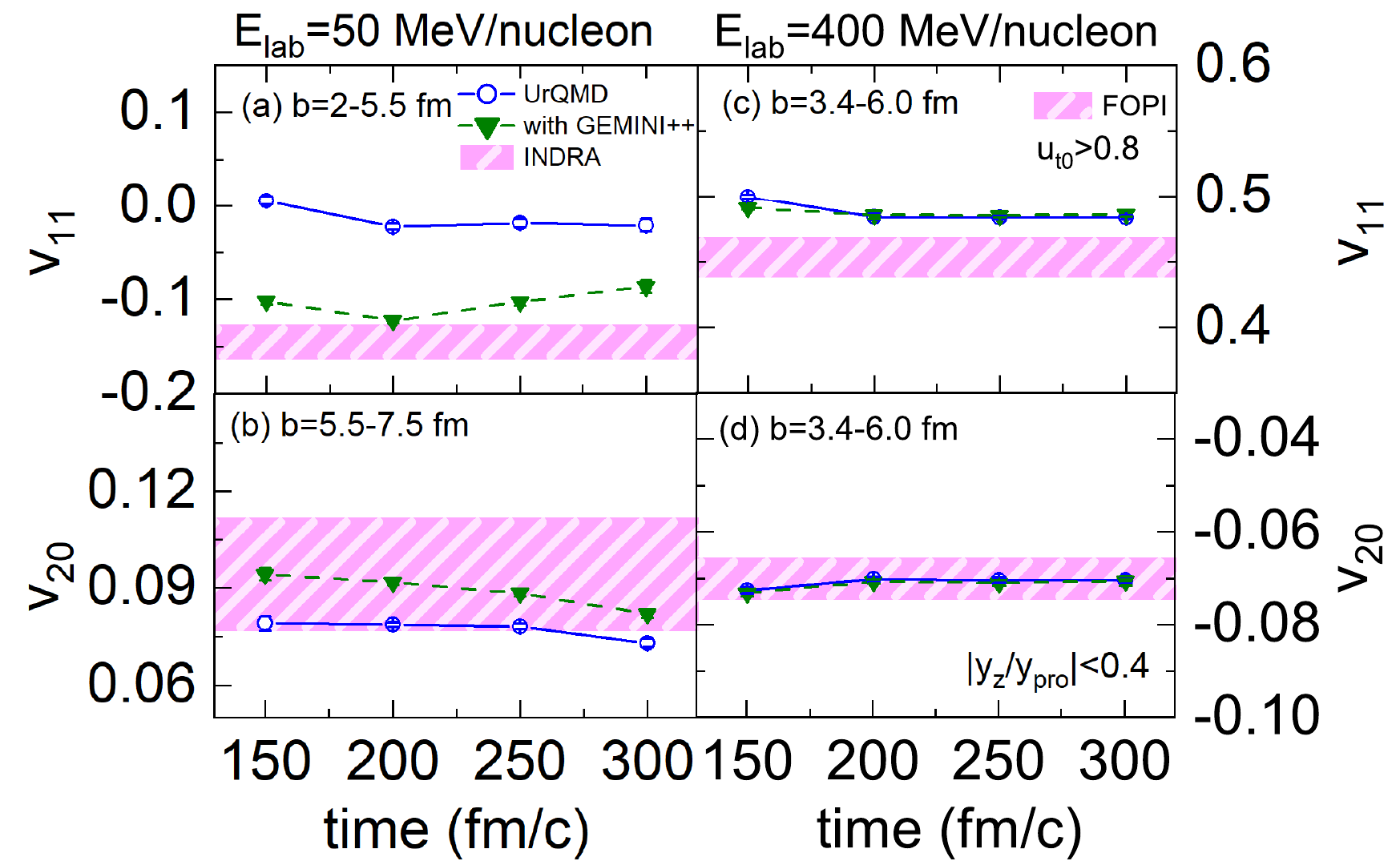}
			\caption{Directed flow slope $v_{11}$ and elliptic flow $v_{20}$ from $^{197}\mathrm{Au}+^{197}\mathrm{Au}$ collisions at $E_{\rm{lab}}$ = 50 MeV/nucleon (for hydrogen isotopes (Z = 1)) and 400 MeV/nucleon (for free protons with $u_{t0}$ > 0.8) with different stopping times. The INDRA and FOPI experimental data taken from Refs.~\cite{Andronic:2006ra,FOPI:2011aa,LeFevre:2016vpp} are indicated by shaded bands}
			\label{fig:3}
		\end{figure}
		
To quantitatively evaluate the influence of sequential decay on the collective flows, as shown in Fig.~\ref{fig:3} the $v_{1}$ slope (top panels) and $v_{2}$ (bottom panels) at mid-rapidity for Z = 1 particles (left panels) and free protons (right panels) are calculated at different stopping times and compared with the experimental data. The slope of directed flow $v_{11}$ and elliptic flow $v_{20}$ at mid-rapidity are extracted by assuming $v_{1}(y_{0})=v_{10}+v_{11}\cdot y_{0}+v_{13}\cdot y_{0}^{3}$ and $v_{2}(y_{0})=v_{20}+v_{22}\cdot y_{0}^{2}+v_{24}\cdot y_{0}^{4}$ in the range of $|y_{0}|$ < 0.4, in the same way as done in the experimental report \cite{Andronic:2006ra,FOPI:2011aa,LeFevre:2016vpp}. The left panels show the results from the simulations at $E_{\rm{lab}}$ = 50 MeV/nucleon, whereas the right panels show the results from the simulations at $E_{\rm{lab}}$ = 400 MeV/nucleon. There is an obvious difference in $v_{11}$ and $v_{20}$ between the simulations with and without considering the decay of the primary fragments at $E_{\rm{lab}}$ = 50 MeV/nucleon. By considering the decay of the primary fragments, the value of $v_{11}$ decreases, while that of $v_{20}$ increases, and both are closer to the experimental data. This can be understood from the fact that both directed and elliptic flows are stronger for large-mass fragments \cite{Andronic:2006ra,FOPI:2011aa,Wang:2013wca}. The decay of fragments
is isotropic within the fragments c.m. in the statistical model. Thus, the light nuclei produced from the decay of primary fragments retain information about the primary fragments. Furthermore, at $E_{\rm{lab}}$ = 50 MeV/nucleon, $v_{11}$ first decreases with time and then saturates (increases) after 200 fm/$c$ in the case without (with) the sequential decay effect. This can be explained by the competition between sequential decay and dynamical effects. At approximately $t$ = 150 fm/$c$, the nucleons are located in a low-density environment, and the final state interactions are attractive; thus, a decrease $v_{11}$ is observed. When considering GEMINI++ at $t$ = 300 fm/$c$, there are only a few large-mass fragments that can emit hydrogen isotopes, and the effect of sequential decay on $v_{11}$ gradually weakens. A comparison of the sequential decay effects at $E_{\rm{lab}}$ = 50 MeV/nucleon with those at $E_{\rm{lab}}$ = 400 MeV/nucleon shows that the sequential decay effects are more pronounced at lower beam energies. This is understandable because more large-mass fragments, which contribute to the production of light clusters, are formed at lower beam energies.

		\begin{figure}[t]
			\centering
			\includegraphics[width=\linewidth]{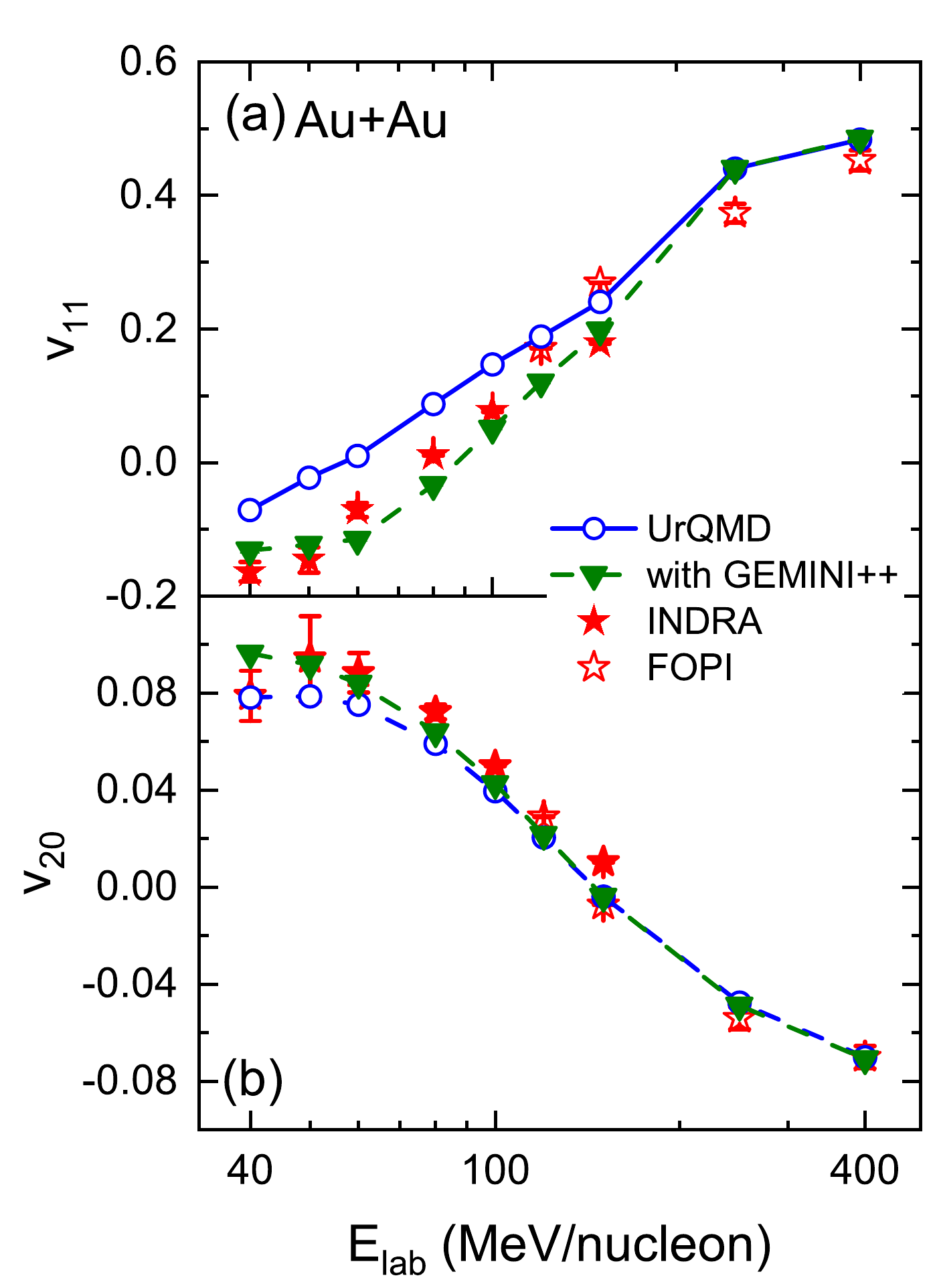}
			\caption{Beam energy dependence of directed flow slope $v_{11}$ (panel a) and elliptic flow $v_{20}$ (panel b) from semi-central $^{197}\mathrm{Au}+^{197}\mathrm{Au}$ collisions. As shown, the solid stars represent the data from INDRA Collaboration \cite{Andronic:2006ra} for Z = 1 particles, and the open stars are the results from FOPI Collaboration \cite{FOPI:2011aa,LeFevre:2016vpp} for free protons with $u_{t0}$ > 0.8 cut. The lines with different symbols are calculated with and without considering the sequential decay. Where error bars are not shown, they fall within the symbols}
			\label{fig:4}
		\end{figure}
		
Fig.~\ref{fig:4} shows the beam energy dependence of $v_{11}$ (top panel) and $v_{20}$ (bottom panel) for semi-central $^{197}\mathrm{Au}+^{197}\mathrm{Au}$ collisions. The value of $v_{11}$ from the simulations without considering sequential decay is higher than that of the experimental data at lower energies. When sequential decay is included, $v_{11}$ is driven down and the $v_{20}$ is pulled up at low energies, which are consistent with the results shown in Fig.~\ref{fig:3}. Again, both $v_{11}$ and $v_{20}$ are considerably closer to the experiment data for all investigated energies. Moreover, the effects of sequential decay on the collective flows decrease with increasing beam energy. With increasing beam energy, the reaction becomes more violent, and the multiplicities of large-mass primary fragments decrease.

		\begin{figure}[t]
			\centering
			\includegraphics[width=\linewidth]{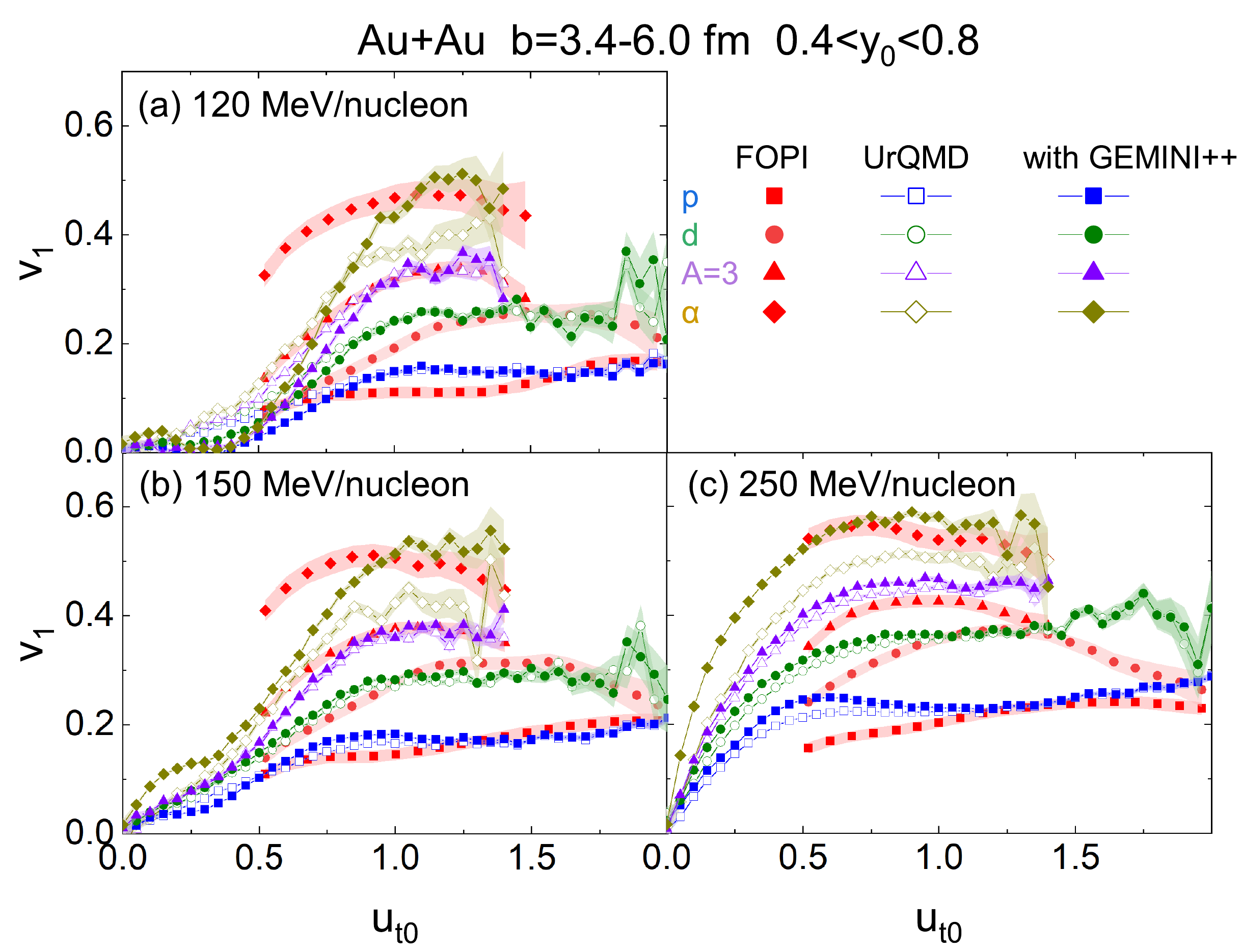}
			\caption{Directed flow $v_{1}$ of protons (squares), deuterons (circles), $A$ = 3 clusters (triangles), and $\alpha$-particles (rhombuses) versus $u_{t0}$ in $^{197}\mathrm{Au}+^{197}\mathrm{Au}$ collisions at $E_{\rm{lab}}$ = 120, 150, 250 MeV/nucleon with $b$ = 3.4-6.0 fm. Results from calculations with and without sequential decay are represented by solid and open symbols, the FOPI experimental data (red symbols) are taken from Ref.~\cite{FOPI:2011aa}. The shaded areas represent the corresponding error bars}
			\label{fig:5}
		\end{figure}

For a systematic investigation of the effects of the sequential decay on the collective flows, the transverse 4-velocities $u_{t0}$ dependence of the $v_{1}$ of light charged particles in semi-central $^{197}\mathrm{Au}+^{197}\mathrm{Au}$ collisions at $E_{\rm{lab}}$ = 120, 150, 250 MeV/nucleon are calculated with and without considering the decay of primary fragments, and compared with the experimental data \cite{FOPI:2011aa}, as shown in Fig.~\ref{fig:5}. The effects of sequential decay on $v_{1}$ ($u_{t0}$) of particles with mass number $A$ < 4 are relatively weak; it only affects $v_{1}$ at low $u_{t0}$. By considering the sequential decay effect, $v_{1}$ of $\alpha$ particles is obviously influenced and can reproduce the experimental data reasonably well at higher $u_{t0}$. This is because after the decay of excited primary fragments is considered, the multiplicity of $\alpha$ particles is enhanced, as shown in Fig.~\ref{fig:1}. The $\alpha$ particles produced by the decay of the excited primary fragments have a large flow effect because they inherit the flow information of the excited primary fragments. The remaining discrepancies in $u_{t0}$-dependent $v_{1}$ of light clusters may be due to simplifications in the initial wave function of particles (nucleons and possible clusters) and quantum effects in two-body collisions, as well as the lack of dynamical production process, which requires further study \cite{Ono:2019jxm}.
		
%%%%%%%%%%%%%%%%%%%%%%%%%%%%%%%%%%%%%%%%%%%%%%%%%%
\subsection{Nuclear stopping power}

		\begin{figure}[t]
			\centering
			\includegraphics[width=\linewidth]{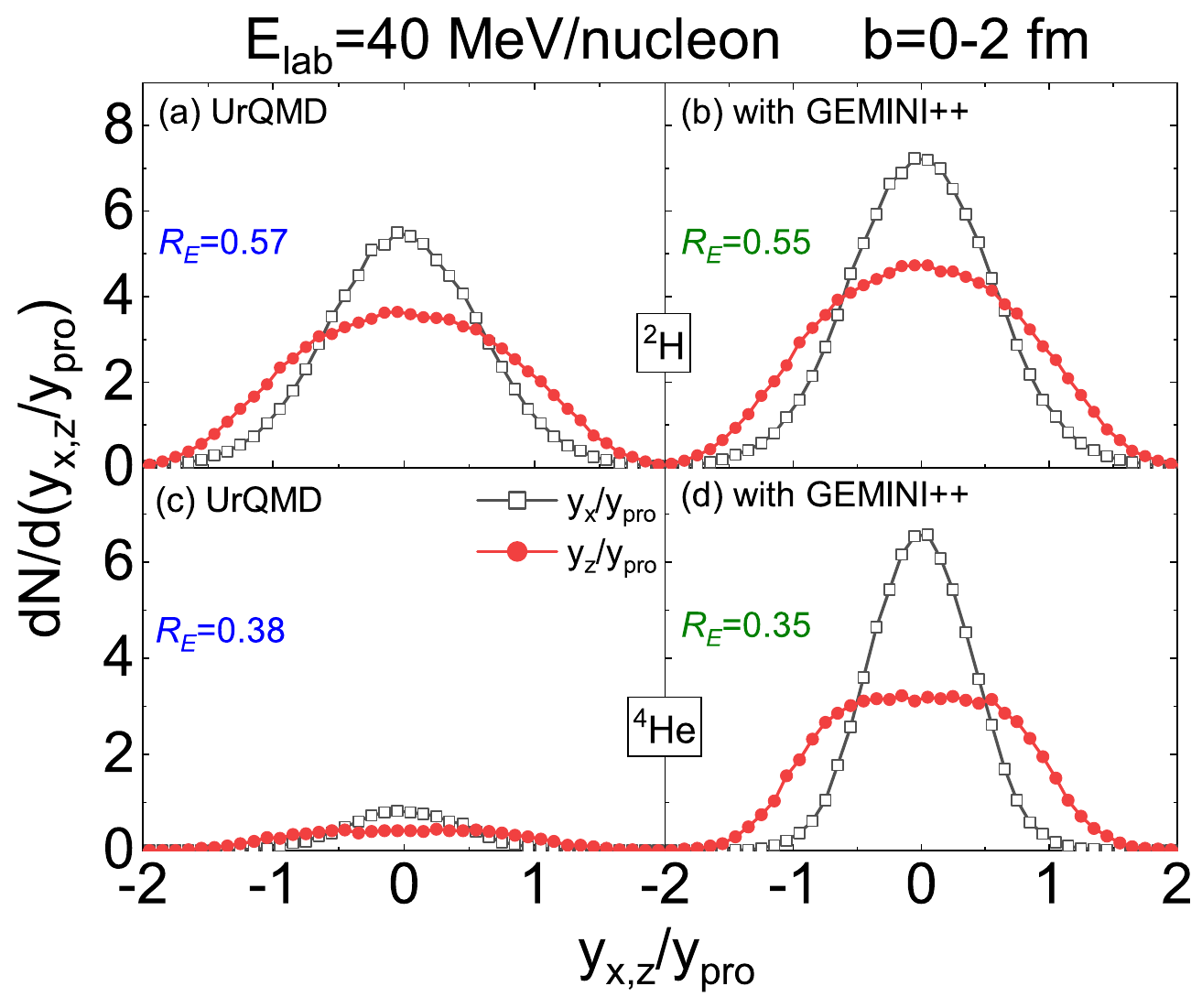}
			\caption{Yield distributions of $^2$H and $^4$He particles as functions of the reduced longitudinal and transverse rapidities from central ($b$ = 0-2 fm) $^{197}\mathrm{Au}+^{197}\mathrm{Au}$ collisions at $E_{\rm{lab}}$ = 40 MeV/nucleon. Results from calculations with and without the sequential decay are shown in the left and right panels. The corresponding values for nuclear stopping power $R_{E}$ are also indicated in each panel}
			\label{fig:6}
		\end{figure}
		
		\begin{figure}[b]
			\centering
			\includegraphics[width=\linewidth]{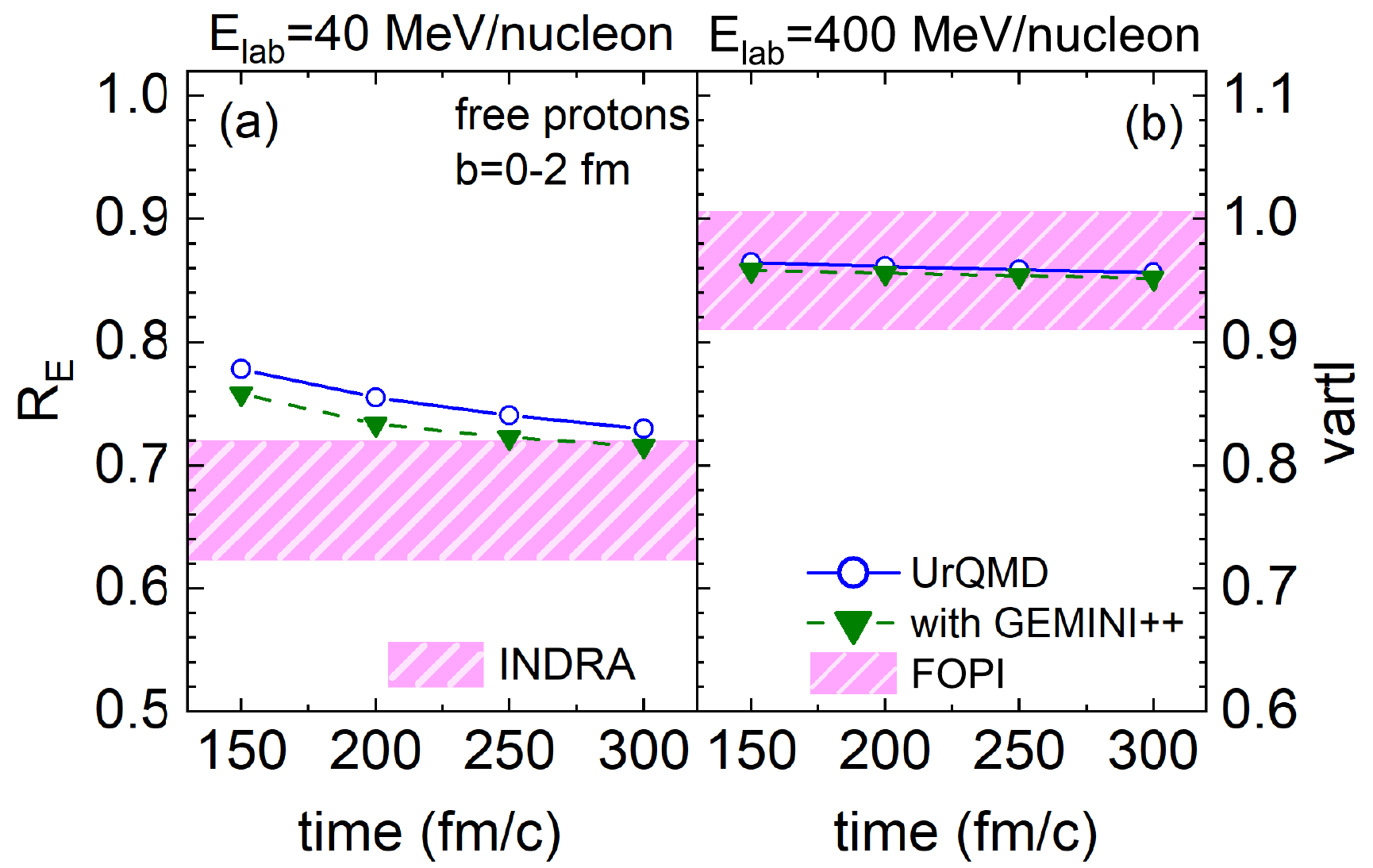}
			\caption{$\mathrm{R_{E}}$ and $vartl$ for free protons from central ($b$ = 0-2 fm) $^{197}\mathrm{Au}+^{197}\mathrm{Au}$ collisions with $E_{\rm{lab}}$ = 40 and 400 MeV/nucleon at different stopping times. The symbols sets are the same as Fig.~\ref{fig:1}. The shaded bands are the corresponding INDRA and FOPI experimental data taken from Refs.~\cite{Lopez:2014dga,FOPI:2010xrt}.}
			\label{fig:7}
		\end{figure}
		
		\begin{figure}[t]
			\centering
			\includegraphics[width=\linewidth]{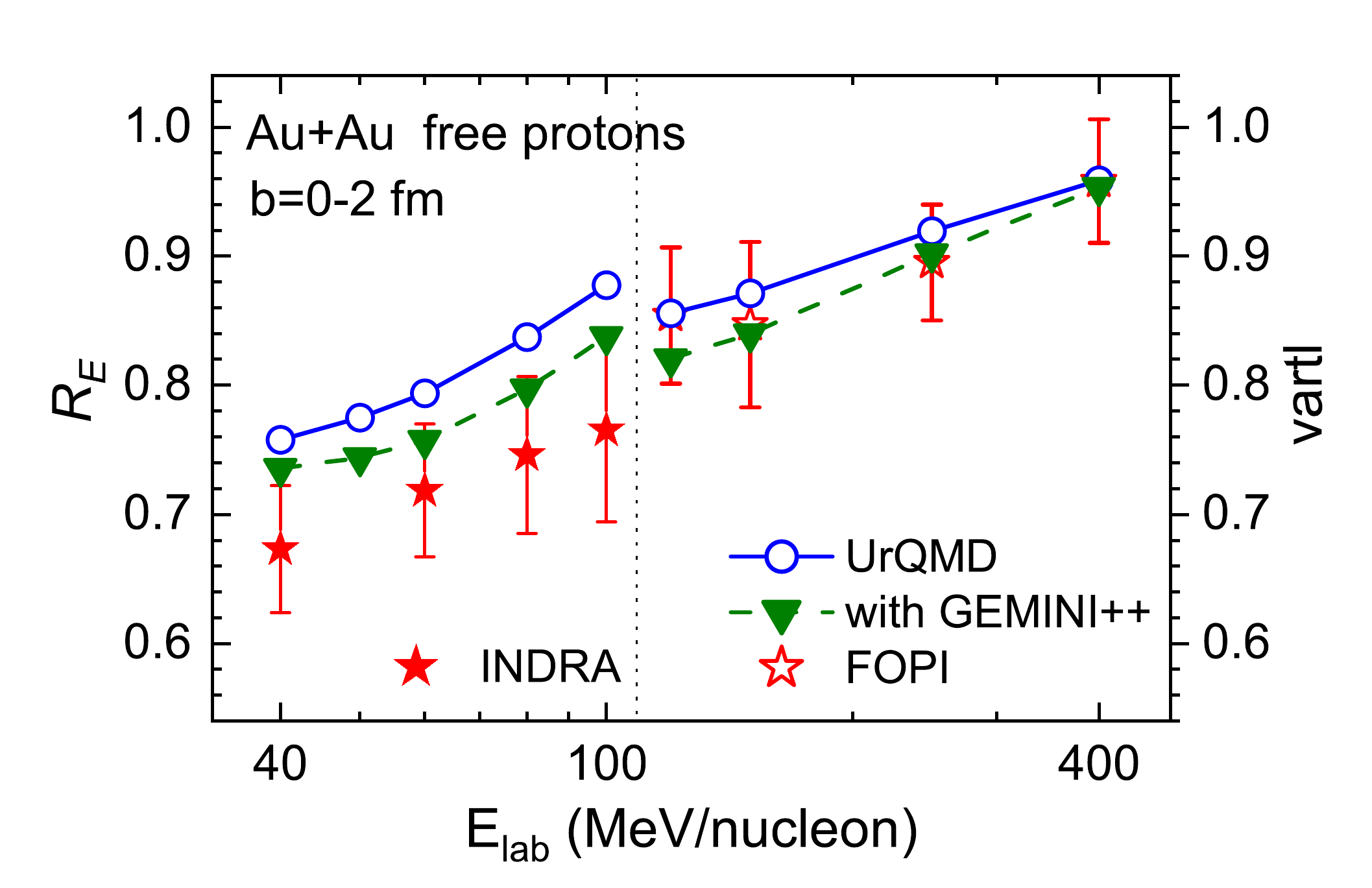}
			\caption{Beam-energy dependence of $R_{E}$ (INDRA) and $vartl$ (FOPI) for free protons from central ($b$ = 0-2 fm) $^{197}\mathrm{Au}+^{197}\mathrm{Au}$ collisions. The symbols sets are the same as Fig.~\ref{fig:5}. Calculations with and without sequential decay are compared with the INDRA and FOPI  experimental data, which are taken from Refs.~\cite{Lopez:2014dga,FOPI:2010xrt}, respectively}
			\label{fig:8}
		\end{figure}
		
The nuclear stopping power characterizes the transparency of the colliding nuclei and can provide insight into the rate of equilibration of the colliding system. Thus, it is meaningful to explore the effects of sequential decay on the nuclear stopping power. The yield distributions of deuterons and $^{4}$He clusters as functions of the reduced longitudinal ($y_{z}/y_{pro}$) and transverse ($y_{x}/y_{pro}$) rapidities for central $^{197}\mathrm{Au}+^{197}\mathrm{Au}$ collisions at $E_{\rm{lab}}$ = 40 MeV/nucleon with and without considering the sequential decay are shown in Fig.~\ref{fig:6}. The value of calculated $R_{E}$ are shown in each panels, and they are slightly decreased; that is, the nuclear stopping power is depressed when considering the sequential decay. Notably, the values of $R_{E}$ are slightly influenced by sequential decay; however, the yield spectrum of each particle is obviously influenced, especially for $^{4}$He clusters. In conjunction with Fig.~\ref{fig:1}, it can be concluded that sequential decay has a strong effect on the particle yield and yield distribution; however, the nuclear stopping power is only weakly influenced because the de-excitation of the primary fragments in the GEMINI++ code is isotropic.
		
The nuclear stopping power $R_{E}$ ($E_{\rm{lab}}$ = 40 MeV/nucleon, left panel) and $vartl$ ($E_{\rm{lab}}$ = 400 MeV/nucleon, right panel) for free protons from central ($b$ = 0-2 fm) $^{197}\mathrm{Au}+^{197}\mathrm{Au}$ collisions calculated at different stopping times are shown in Fig.~\ref{fig:7}. Similar to the collective flows shown in Fig.~\ref{fig:3}, the effects of sequential decay on the nuclear stopping power at lower energies are more obvious than those at higher energies. In all cases, the results were saturated and were close to or covered by the experimental data above 200 fm/$c$.
	
Fig.~\ref{fig:8} shows the degree of nuclear stopping ($R_{E}$ or $vartl$) in the central $^{197}\mathrm{Au}+^{197}\mathrm{Au}$ collisions as a function of the beam energy. It is clear that both $R_{E}$ and $vartl$ decrease and are much closer to the experimental data when the GEMINI++ code is applied, and the difference in the values of the nuclear stopping power gradually disappears with increasing beam energy. With the same impact parameter (centrality), more violent reactions occurred at higher beam energies, and less heavily excited primary fragments were produced. The difference with free protons is that they are produced from the decay of heavy excited primary fragments, which usually have a weak stopping power (smaller values of $R_{E}$ and $vartl$) \cite{FOPI:2004orn,FOPI:2010xrt}.  As a result, the nuclear stopping power was slightly reduced by including the GEMINI++ code.
%%%%%%%%%%%%%%%%%%%%%%%%%%%%%%%%%%%%%%%%%%%%%%%%
\section{Summary and outlook }
\label{sec:4}
In summary, the effects of the sequential decay of excited primary fragments on the rapidity distribution, collective flows of light nuclei, and nuclear stopping power in Au+Au collisions at intermediate energies were investigated. Primary fragments were produced using the UrQMD model, and sequential decay of the excited primary fragments was performed using the GEMINI++ code. It was observed that the sequential decay of the excited primary fragments have an obvious influence on the rapidity distribution and collective flows but relatively weakly affect the stopping power. This is because of the memory effect; that is, the light particles produced from the sequential decay of the excited primary fragments inherit the collective properties of the excited primary fragments. Furthermore, the sequential decay effects gradually decreased with increasing beam energy because fewer large-mass fragments that can emit light particles were produced at higher beam energies. More importantly, the ability to reproduce the relevant experimental data at the investigated energies was improved to a certain extent by the inclusion of sequential decay effects.
            
These results are meaningful for extracting the properties of nuclear matter by comparing transport model calculations with experimental data. The effects of secondary sequential decay on the observables, such as the $N/Z$ ratios and elliptic flow ratios of neutrons vs. hydrogen isotopes, will be discussed in future publications.
%%%%%%%%%%%%%%%%%%%%%%%%%%%%%%%%%%%%%%%%%%%%%%%%
\section*{Acknowledgement}
The authors acknowledge support by computing server C3S2 at the Huzhou University. This work is partly supported by the National Natural Science Foundation of China (Nos.U2032145 and 11875125) and the National Key Research and Development Program of China under Grant No. 2020YFE0202002. 

\bibliography{manuscripts}
\bibliographystyle{elsarticle-num}

\end{document}